\documentclass[conference]{IEEEtran}
\IEEEoverridecommandlockouts
\usepackage{cite}
\usepackage{amsmath,amssymb,amsfonts}
\usepackage{graphicx}
\usepackage{textcomp}
\usepackage{xcolor}
\usepackage{booktabs}
\usepackage{multirow}
\usepackage{url}
\usepackage{stfloats}
\usepackage[htt]{hyphenat}
\def\BibTeX{{\rm B\kern-.05em{\sc i\kern-.025em b}\kern-.08em
    T\kern-.1667em\lower.7ex\hbox{E}\kern-.125emX}}

\begin{document}

\title{Vendor-Conditioned Contrastive Learning for Predicting Organizational Cyber Threat Targets}

\author{\IEEEauthorblockN{Benjamin Ampel}
\IEEEauthorblockA{\textit{Department of Computer Science} \\
\textit{Georgia State University}\\
Atlanta, Georgia, USA \\
bampel@gsu.edu}
}

\maketitle

\begin{abstract}
Cyberattacks cause billions of dollars in damage annually, with malicious hackers often sharing exploit code and techniques on underground forums. Identifying which organizations are targeted by these exploits is critical for proactive Cyber Threat Intelligence (CTI). To address that gap, we propose Temporal Representation and Classification of Exploits (TRACE), a vendor-conditioned contrastive learning framework built on CySecBERT that jointly optimizes organizational target classification and vendor-coherent representations while evaluating robustness under temporal distribution shift. Unlike prior work limited to small, single-source datasets, we leverage a large-scale, multi-source corpus spanning 9 exploit databases and hacker forums, comprising 352,866 posts collected over three decades, yielding a 129,126-sample dataset across seven organizational categories. In the temporal out-of-distribution evaluation, TRACE achieves macro F1=97.00\%, substantially outperforming 17 benchmark classical ML methods, deep learning with GloVe/FastText embeddings, and pretrained transformer models.
\end{abstract}


\section{Introduction}

\begin{figure}[htbp]
\centerline{\includegraphics[width=\columnwidth]{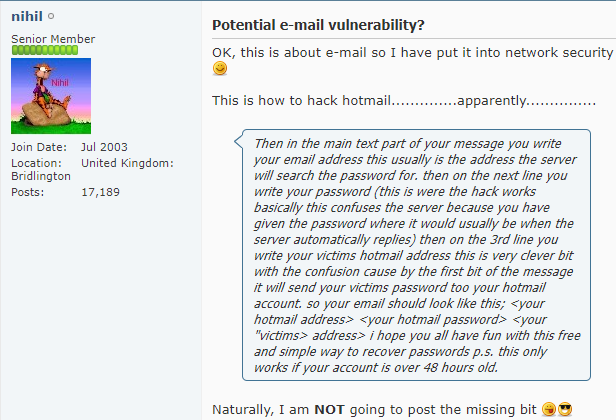}}
\caption{Example hacker forum post discussing an exploit technique targeting a specific organization's email service.}
\label{fig:forum_post}
\end{figure}

Organizations, governments, and individuals face an ever-increasing volume of cyberattacks, with global cybercrime damages estimated at \$10.5 trillion in 2025 and projected to reach \$15.6 trillion annually by 2029~\cite{cybercrime_costs}. To combat this threat, organizations invest heavily in Cyber Threat Intelligence (CTI, the collection, analysis, and dissemination of information about cyber threats)~\cite{hughes2024csur}. A key source of proactive CTI lies in hacker forums, where malicious actors congregate to share exploit source code, discuss attack techniques, and trade tools targeting specific organizations~\cite{samtani2020}. An example of such a forum post is shown in Figure~\ref{fig:forum_post}.

Prior work on hacker forum exploit analysis has focused primarily on exploit identification~\cite{ampel2020labeling}, MITRE ATT\&CK mapping~\cite{ampel2023attck}, and cross-lingual threat detection~\cite{ebrahimi2020}. However, a critical gap remains. Which organizations are being targeted by these exploits? Understanding organizational targeting enables companies and sector-specific agencies to prioritize defenses against threats tailored to their industry.

This gap is especially consequential because hacker communities are not static. As new vendors, products, vulnerabilities, and attack techniques emerge, the language used to describe exploits changes as well. A model trained on historical forum posts may therefore learn patterns that become less reliable when applied to future posts. Organizational target prediction must therefore be evaluated under temporal drift, where the vocabulary, targets, and exploit patterns observed during training may differ from those encountered during deployment.

To address this gap, in this paper, we propose Temporal Representation and Classification of Exploits (TRACE), an organizational detection framework with two key contributions:

\begin{enumerate}
    \item \textbf{Large-scale multi-source corpus}: A dataset of labeled 129,126-samples labeled dataset to enable downstream tasks; and
    \item \textbf{Vendor-conditioned contrastive learning model}: A CySecBERT-based architecture with a supervised contrastive objective over vendor groups, achieving F1=97.00\% on temporal OOD evaluation.
\end{enumerate}

\section{Related Work}

\subsection{Hacker Community Analytics}
A survey~\cite{hughes2024csur} systematizes the methodologies for studying cybercrime communities, establishing hacker forums as a primary data source for proactive CTI~\cite{samtani2020}. In this space, exploit analysis has progressed rapidly. Literature explored deep transfer learning for automated exploit labeling from forum posts~\cite{ampel2020labeling} and extended exploit detection to non-English forums via adversarial cross-lingual knowledge transfer~\cite{ebrahimi2020}. Subsequent work mapped exploit code to the MITRE ATT\&CK framework using multi-label transformers~\cite{ampel2023attck} and profiled the evolution of exploit-sharing hackers through graph embeddings~\cite{otto2021evolution}. Finally, NER-based methods have extracted protected health information from multilingual hacker communities~\cite{dacosta2026phi}. 

Importantly, a systematic study of dark web language demonstrated hacker communities use unique vocabulary that general-purpose Natural Language Processing (NLP) models struggle to generalize to~\cite{jin2022darkweb}. Related literature attempted to improve NLP model generalizability through GloVe~\cite{ampel2024creating}, distilling contextual hacker community embeddings~\cite{ampel2021hack2vec}, and cyberesecurity domain-specific language models~\cite{bayer2024cysecbert, aghaei2022securebert, park2023ctibert}. These models consistently achieve state-of-the-art results on cybersecurity NLP tasks, confirming that cybersecurity domain-specific pretraining is essential for hacker community analytics. 

However, prior work primarily classifies exploits by type or tactic rather than organizational target. Thus, while cybersecurity-specific pretraining improves lexical and contextual coverage, it does not explicitly exploit the vendor-level structure inherent in exploit disclosures, nor does it establish whether organizational target classifiers remain reliable under forward temporal shift. This motivates our vendor-conditioned contrastive fine-tuning objective.

\subsection{Contrastive Learning and Temporal Robustness}
Contrastive learning has emerged as a powerful paradigm for learning discriminative representations. Literature generalized the self-supervised contrastive framework of SimCLR~\cite{chen2020simclr} to the fully-supervised setting~\cite{khosla2020supcon}. This new Supervised Contrastive (SupCon) loss leverages label information to pull same-class embeddings together while pushing different-class embeddings apart~\cite{khosla2020supcon}. This approach outperformed standard cross-entropy on image classification benchmarks. Subsequent literatuer adapted SupCon learning to pretrained language model fine-tuning, demonstrating improved generalization on text classification tasks (including sentiment analysis and natural language inference)~\cite{gunel2021scl_nlp}. It is often observed that contrastive objectives yield state-of-the-art sentence embeddings for downstream NLP tasks~\cite{gao2021simcse}.

However, a growing body of work has documented the challenge of temporal distribution shift in NLP. For example, research found an average 20\% performance degradation when NLP models are evaluated on temporally out-of-distribution (OOD) data~\cite{yao2022wildtime}. This pattern held in several settings, finding that temporal drift erodes classifier performance even when the underlying task remains identical~\cite{guo2023temporal}. 

In the temporal domain, contrastive methods have been applied to learn representations that capture temporal structure~\cite{tonekaboni2021tnc, zhang2022tfc}. For example, Temporal Neighborhood Coding adapts temporal proximity as a self-supervised signal for time-series representation learning, while Zhang et al.~\cite{zhang2022tfc} introduced time-frequency consistency as a contrastive objective for time-series pretraining. However, these approaches are designed primarily for time-series data and do not address the representational challenge posed by cybersecurity text, where exploit posts contain vendor names, product identifiers, version strings, vulnerability references, and code fragments that evolve over time. Existing contrastive approaches also do not specify how auxiliary metadata in exploit disclosures can be used to learn representations that remain discriminative for organizational target prediction under temporal distribution shift. 

\section{Research Design}
Our research design is organized around three challenges in organizational target prediction from hacker community data. The task requires a large and diverse exploit corpus, a high-precision procedure for mapping vendor and product references to organizational categories, and an evaluation protocol that reflects temporal drift in real deployments. Accordingly, we construct a multi-source corpus, extract organizational targets through a curated gazetteer, split the data forward in time, and evaluate TRACE against classical machine learning (ML), deep learning (DL), and pretrained transformer baselines.

\subsection{Dataset}
Our design requires that we construct a corpus that captures exploit discussions across heterogeneous CTI-relevant source layers. To create this dataset, we leveraged the HackerSignal dataset (comprised of 35 public sources and 8.7 million posts collected over three decades)~\cite{ampel2026hackersignal}. We provide the overall summary of HackerSignal in Table~\ref{tab:corpus}.

\begin{table}[htbp]
\caption{Corpus Summary by Source Layer}
\label{tab:corpus}
\begin{center}
\begin{tabular}{lrrr}
\toprule
\textbf{Source Layer} & \textbf{Sources} & \textbf{Posts} & \textbf{Date Range} \\
\midrule
Hacker Forums & 21 & 4,811,505 & 2001--2026 \\
Exploit Databases & 5 & 3,048,765 & 1990--2026 \\
Vuln.\ Advisories & 7 & 385,039 & 2002--2026 \\
Darknet Markets & 1 & 488,989 & 2014--2015 \\
Fix Commits & 1 & 8,362 & 1999--2022 \\
\midrule
\textbf{Total} & \textbf{35} & \textbf{8,742,660} & \textbf{1990--2026} \\
\bottomrule
\end{tabular}
\end{center}
\end{table}

For organizational target prediction, we opted to focus on the English exploit-rich sources most likely to reference targeted organizations: exploit databases (e.g., ExploitDB, 0day.today), bug bounty platforms (e.g., HackerOne), and hacker forums (e.g., HackForums, Go4Expert). This HackerSignal subset accounts for  352,866 posts for entity extraction. Exploit titles and descriptions in these sources often follow predictable patterns (e.g., \textit{VendorName ProductName Version -- VulnType}), enabling precise vendor extraction via a gazetteer (a curated dictionary of known entity names) of 150+ vendor and product name patterns organized by industry category. Each post is matched against the title and the opening 500 characters of description text. This gazetteer-based approach yields higher precision than general-purpose NER on noisy exploit code, where register names and technical terms are frequently misclassified as organizations.

Extracted vendors are normalized and mapped to industry categories (e.g., Networking, Gaming). Categories with fewer than 100 samples were removed. Of the 352,866 posts processed, 129,126 (36.6\%) contained identifiable organizational targets, forming our gold-standard dataset of vendors mentioned in known exploits (Table~\ref{tab:goldstandard}).

\begin{table}[htbp]
\caption{Gold-Standard Dataset by Organization Category}
\label{tab:goldstandard}
\begin{center}
\begin{tabular}{lrr}
\toprule
\textbf{Category} & \textbf{Count} & \textbf{Pct.} \\
\midrule
CMS / Open Source & 44,791 & 34.7\% \\
Enterprise Software & 44,460 & 34.4\% \\
Web Platform & 30,294 & 23.5\% \\
Networking / IoT & 4,672 & 3.6\% \\
Mobile & 3,272 & 2.5\% \\
Gaming & 1,027 & 0.8\% \\
Database & 610 & 0.5\% \\
\midrule
\textbf{Total} & \textbf{129,126} & \textbf{100\%} \\
\bottomrule
\end{tabular}
\end{center}
\end{table}

\subsection{Preprocessing and Temporal Splitting}
Each post's raw text is cleaned by removing non-UTF-8 characters and normalizing whitespace. The exploit source code is retained because it contains valuable organizational references (e.g., target hostnames, library imports). We split the data temporally to avoid future leakage. Posts before 2022 form the training set (125,963 samples), 2022--2023 posts form the validation set, and 2024+ posts form the test set. Posts without timestamps are discarded. This temporal out-of-distribution (OOD) protocol requires models to generalize to future and potentially unseen exploit patterns. This data as used as input into our TRACE framework

\subsection{Temporal Representation and Classification of Exploits (TRACE)}

We propose TRACE, a vendor-conditioned contrastive learning framework that optimizes organizational target classification while encouraging vendor-coherent representations. The architecture consists of three components, each motivated by the characteristics of exploit text classification under temporal distribution shift.

\textbf{1. Shared encoder.} We build on CySecBERT~\cite{bayer2024cysecbert}, a BERT-base encoder pretrained on cybersecurity corpora. Domain-adapted pretraining is useful for exploit text that contains specialized vocabulary such as CVE identifiers, shell commands, obfuscated payloads, product names, and version strings. The encoder processes tokenized exploit text and produces a contextualized \texttt{[CLS]} representation $\mathbf{h} \in \mathbb{R}^{768}$.

\textbf{2. Classification head.} A single linear layer $W_c \in \mathbb{R}^{768 \times 7}$ maps $\mathbf{h}$ to class logits, trained with standard cross-entropy loss $\mathcal{L}_{CE}$. We keep the TRACE classification head minimal so that representational capacity remains concentrated in the shared encoder, where the contrastive objective can shape the learned representation.

\textbf{3. Contrastive projection head.} A two-layer MLP ($768 \rightarrow 768 \rightarrow 128$ with ReLU) projects $\mathbf{h}$ into a contrastive embedding space, followed by $\ell_2$ normalization: $\mathbf{z} = \mathrm{normalize}(g(\mathbf{h})) \in \mathbb{R}^{128}$. Following SupCon learning~\cite{khosla2020supcon}, the contrastive objective pulls positive examples together and pushes negative examples apart. In TRACE, positives are defined by vendor identity rather than by organizational category labels or temporal bins. A SupCon loss pulls posts from the same \textit{vendor} together while pushing posts from different vendors apart:
\begin{equation}
\mathcal{L}_{SC} = -\frac{1}{|B|}\sum_{i \in B} \frac{1}{|P_i|}\sum_{p \in P_i} \log \frac{\exp(\mathbf{z}_i \cdot \mathbf{z}_p / \tau)}{\sum_{j \neq i} \exp(\mathbf{z}_i \cdot \mathbf{z}_j / \tau)}
\end{equation}
where $B$ is a mini-batch, $P_i = \{p \in B : \mathrm{vendor}(p) = \mathrm{vendor}(i), p \neq i\}$ is the set of same-vendor positives, and $\tau = 0.07$ is the temperature.

The rationale for vendor rather than label-based grouping is twofold. First, exploit posts associated with the same vendor or product family often share vocabulary, code patterns, vulnerability identifiers, and version-specific language that may cut across broader organizational categories. Second, vendor grouping provides a more granular auxiliary supervision signal than the seven-category target label. The classification loss separates broad organizational categories, while the contrastive loss encourages the embedding space to preserve vendor-level structure.

\begin{figure}[htbp]
\centerline{\includegraphics[width=\columnwidth]{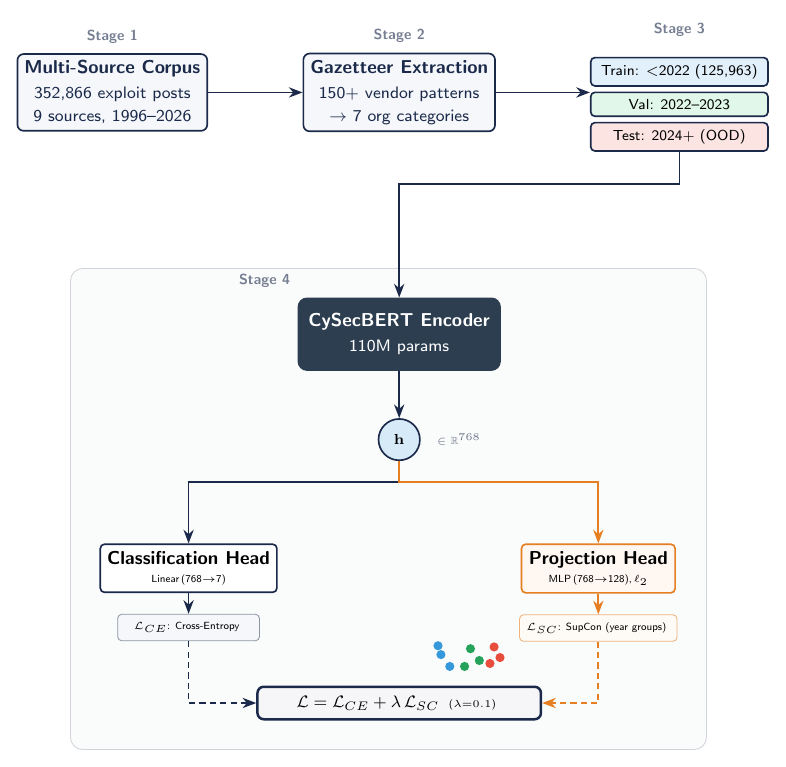}}
\caption{TRACE research design: multi-source corpus $\rightarrow$ gazetteer extraction $\rightarrow$ temporal split $\rightarrow$ dual-head CySecBERT model with vendor-conditioned contrastive learning (novelty in orange) $\rightarrow$ temporal OOD evaluation.}
\label{fig:architecture}
\end{figure}

The total training objective combines both losses:
\begin{equation}
\mathcal{L} = \mathcal{L}_{CE} + \lambda \mathcal{L}_{SC}, \quad \lambda = 0.1
\end{equation}
where $\lambda = 0.1$ down-weights the contrastive term so that classification remains the primary objective. At inference, the classification head produces predictions. The projection head is used only to shape representations during training. Training uses AdamW~\cite{loshchilov2019adamw} with learning rate $2 \times 10^{-5}$. In our implementation, TRACE is fine-tuned for three epochs and is evaluated using the temporal out-of-distribution test protocol. Figure~\ref{fig:architecture} illustrates the full TRACE pipeline from corpus to evaluation.

\subsection{Benchmark Models}
We compare TRACE against benchmark models spanning three paradigms. All models are trained on the full 125,963 training samples and evaluated on the temporal OOD test set.

\begin{itemize}
\item \textbf{Classical ML} (4 models). TF-IDF features (unigrams + bigrams, sublinear TF) fed to Na\"ive Bayes, Logistic Regression, Decision Tree, and LinearSVC.

\item \textbf{Deep Learning} (14 models). Seven architectures (RNN, GRU, LSTM, BiLSTM, CNN-LSTM, BiLSTM with attention, and a shallow Transformer encoder~\cite{vaswani2017attention}) each trained with two embedding initializations. First, a generic GloVe 6B 300d~\cite{pennington2014glove}. Second, a FastText 300d~\cite{bojanowski2017fasttext} trained on exclusively on our exploit training corpus. All use 256-dimensional hidden states, 0.3 dropout, the Adam optimizer (lr=$10^{-3}$), early stopping (patience=2), and train for up to 20 epochs.

\item \textbf{Pretrained Transformers} (LLMs). BERT~\cite{devlin2019bert}, RoBERTa~\cite{liu2019roberta}, ModernBERT~\cite{warner2024modernbert}, and CySecBERT~\cite{bayer2024cysecbert}, each fine-tuned with early stopping for up to 10 epochs.

\end{itemize}

\section{Experiments and Results}

\subsection{Evaluation Metrics}
We report Accuracy, macro F1-Score, macro Precision, and macro Recall on the temporal OOD test set. Models were assessed via paired bootstrap tests (1,000 resamples) to determine the significance of TRACE relative to each benchmark.

\subsection{Results}

Table~\ref{tab:results} presents results across all model groups.

\begin{table}[htbp]
\caption{Experiment Results: Temporal OOD Evaluation (\%)}
\label{tab:results}
\begin{center}
\small
\begin{tabular}{llcccc}
\toprule
\textbf{Group} & \textbf{Model} & \textbf{Acc.} & \textbf{F1} & \textbf{Prec.} & \textbf{Rec.} \\
\midrule
\multirow{4}{*}{\rotatebox{90}{\small Class.\ ML}}
& Na\"ive Bayes$^{***}$ & 79.52 & 40.01 & 55.26 & 39.45 \\
& Decision Tree$^{***}$ & 88.55 & 62.43 & 70.26 & 58.58 \\
& Logistic Reg.$^{***}$ & 91.36 & 73.42 & 74.16 & 73.19 \\
& SVM (Linear)$^{***}$ & 92.65 & 75.69 & 75.50 & 76.26 \\
\midrule
\multirow{7}{*}{\rotatebox{90}{\small DL + GloVe}}
& RNN$^{***}$ & 22.42 & 10.42 & 11.87 & 18.54 \\
& LSTM$^{***}$ & 71.75 & 57.33 & 60.52 & 57.95 \\
& BiLSTM$^{***}$ & 79.42 & 63.11 & 64.02 & 64.39 \\
& GRU$^{***}$ & 81.31 & 65.18 & 65.32 & 66.83 \\
& CNN-LSTM$^{***}$ & 77.36 & 68.20 & 69.64 & 70.88 \\
& BiLSTM-Attn$^{***}$ & 82.60 & 71.65 & 71.20 & 74.51 \\
& Transformer$^{***}$ & 77.42 & 72.38 & 82.31 & 69.36 \\
\midrule
\multirow{7}{*}{\rotatebox{90}{\small DL + FastText}}
& RNN$^{***}$ & 65.64 & 22.46 & 22.16 & 23.08 \\
& LSTM$^{***}$ & 90.87 & 61.45 & 65.78 & 59.73 \\
& CNN-LSTM$^{***}$ & 94.65 & 72.33 & 72.87 & 71.89 \\
& Transformer$^{***}$ & 96.33 & 77.18 & 79.84 & 75.74 \\
& BiLSTM-Attn$^{***}$ & 96.43 & 80.54 & 80.35 & 80.80 \\
& GRU$^{***}$ & 94.92 & 81.18 & 88.06 & 77.34 \\
& BiLSTM$^{***}$ & 93.25 & 81.84 & 87.16 & 79.85 \\
\midrule
\multirow{4}{*}{\rotatebox{90}{\small LLM}}
& BERT$^{***}$ & 97.78 & 89.29 & 90.49 & 91.91 \\
& RoBERTa$^{***}$ & 96.49 & 84.36 & 87.84 & 82.77 \\
& ModernBERT$^{***}$ & 97.73 & 88.11 & 89.93 & 86.54 \\
& CySecBERT$^{**}$ & 97.51 & 94.45 & 98.35 & 92.32 \\
\midrule
& \textbf{TRACE (Ours)} & \textbf{98.65} & \textbf{97.00} & \textbf{96.63} & \textbf{97.38} \\
\bottomrule
\multicolumn{6}{l}{\scriptsize Paired bootstrap vs.\ TRACE (1,000 resamples):}\\
\multicolumn{6}{l}{\scriptsize $^{***}p{<}0.001$,\; $^{**}p{<}0.01$,\; $^{*}p{<}0.05$,\; $^{\dagger}p{<}0.10$}\\
\end{tabular}
\end{center}
\end{table}

Our results yielded several notable insights. First, TRACE achieved the strongest performance (F1=97.00\%, Acc=98.65\%). The vendor-conditioned contrastive objective encourages the CySecBERT encoder to learn vendor-coherent representations that generalize to future exploit patterns. Second, domain-trained embeddings matter more than architecture for deep learning baselines. FastText embeddings (trained on our exploit corpus) consistently outperformed pretrained GloVe across all architectures (e.g., BiLSTM: F1 81.84\% vs.\ 63.11\%). FastText's subword representations handle the out-of-vocabulary tokens prevalent in exploit text (CVE identifiers, tool names, obfuscated strings) and its corpus-specific training captures the distributional semantics of hacker discourse. This suggests that domain-specific data is a powerful factor across all model families.

Third, among classical ML models, SVM with TF-IDF (F1=75.69\%) outperformed all GloVe-based DL models, confirming that sparse n-gram features capture vendor-specific terminology that generalizes across time. The FastText-based models and pretrained transformers surpass SVM, demonstrating that distributed representations trained on large corpora can provide superior generalization when vocabulary coverage is adequate. Fourth, the vanilla RNN failed to converge reliably with either embedding (F1=10.42\% GloVe, 22.46\% FastText), confirming the known vanishing gradient problem on long exploit text sequences.

Finally, domain-specific pretraining is decisively necessary. CySecBERT (F1=94.45\%) outperforms BERT (89.29\%), ModernBERT (88.11\%), and RoBERTa (84.36\%). CySecBERT's vocabulary, trained on cybersecurity corpora, preserves exploit-specific tokens (CVE IDs, function names, shell commands) as meaningful units rather than fragmenting them into uninformative subwords. Among general-purpose models, RoBERTa's BPE tokenizer fragments these tokens most aggressively, while ModernBERT's architectural improvements (rotary embeddings, flash attention) provide marginal gains but do not overcome the vocabulary bottleneck. TRACE's vendor-conditioned contrastive objective further improves upon standalone CySecBERT fine-tuning (F1: 97.00\% vs.\ 94.45\%, a 2.55\% gain), demonstrating that the vendor-conditioned SupCon loss provides a substantial complementary signal beyond domain-adapted initialization.

\subsection{Per-Category Analysis}

To understand where TRACE's vendor-conditioned contrastive objective improves over standalone CySecBERT fine-tuning, Table~\ref{tab:perclass} compares per-category F1 scores on the temporal OOD test set.

\begin{table}[htbp]
\caption{Per-Category F1 Score: CySecBERT vs.\ TRACE}
\label{tab:perclass}
\begin{center}
\small
\setlength{\tabcolsep}{5pt}
\renewcommand{\arraystretch}{1.15}
\begin{tabular}{@{}l r r r r@{}}
\toprule
\textbf{Category} & \textbf{Test} & \textbf{CySecBERT} & \textbf{TRACE} & \textbf{$\Delta$F1} \\
\midrule
CMS / Open Src. & 272 & 96.09 & 98.55 & +2.46 \\
Enterprise Softw. & 350 & 97.88 & 98.01 & +0.13 \\
Web Platform & 1,108 & 98.23 & 99.32 & +1.09 \\
Network / IoT & 48 & 97.87 & 93.75 & $-$4.12 \\
Mobile & 41 & 98.77 & 97.56 & $-$1.21 \\
Gaming & 30 & 72.34 & 91.80 & +19.46 \\
Database & 2 & 100.00 & 100.00 & 0.00 \\
\midrule
\textbf{Macro Avg} & \textbf{1,851} & \textbf{94.45} & \textbf{97.00} & \textbf{+2.55} \\
\bottomrule
\end{tabular}
\end{center}
\end{table}

TRACE outperforms CySecBERT across 5 of 7 categories, with the largest gain in Gaming ($+19.46\%$ F1). CySecBERT struggles in this low-support category ($F1=72.34\%$, 30 test samples), misclassifying nearly half of the instances. TRACE's vendor-conditioned contrastive objective raised Gaming F1 to 91.80\% by learning vendor-coherent representations that better separate rare classes under distribution shift. TRACE also improves CMS/Open Source ($+2.46\%$) and Web Platform ($+1.09\%$). The slight regressions on Networking/IoT ($-4.12\%$) and Mobile ($-1.21\%$) reflect trade-offs in the macro-averaged objective, though both categories retain strong absolute performance ($>93\%$ F1).

\begin{figure*}[t]
\centerline{\includegraphics[width=\textwidth]{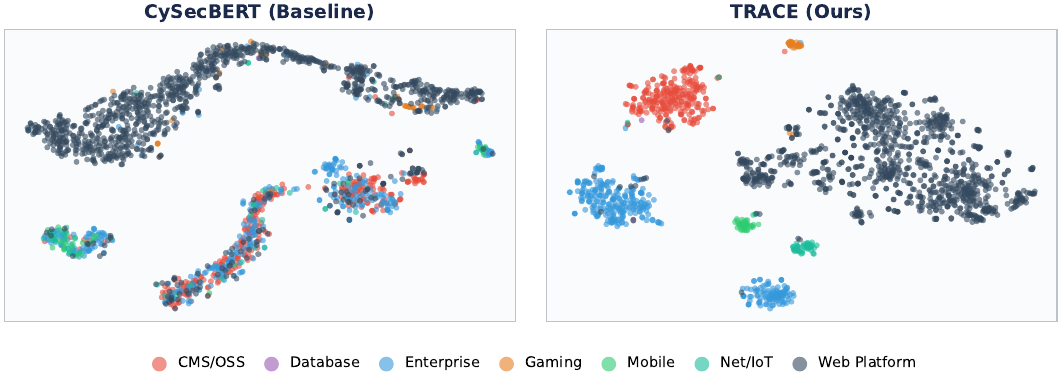}}
\caption{t-SNE projections of test-set [CLS] representations. CySecBERT baseline embeddings (left) show substantial category overlap. TRACE embeddings (right) form well-separated clusters.}
\label{fig:tsne}
\end{figure*}

TRACE makes only 25 errors out of 1,851 test samples (98.65\% accuracy). Examining the misclassifications reveals that most errors occur at genuine category boundaries. For example, the Ruckus IoT Controller (Networking/IoT) is predicted to be CMS/Open Source. Ruckus devices run embedded Linux with web management interfaces, sharing vocabulary with open-source CMS exploits. Similarly, GitLab (Web Platform) is classified as a CMS/Open Source, reflecting its dual identity as both a web application and an open-source project. VMware Cloud Director (Web Platform) is predicted as Enterprise Software, a reasonable confusion given VMware's enterprise positioning. These boundary cases suggest that the 7-category taxonomy, while practically useful, imposes hard boundaries on organizations that straddle multiple categories. Figure~\ref{fig:tsne} visualizes this effect. CySecBERT's representations exhibit substantial inter-category overlap, whereas TRACE produces more tightly clustered, well-separated embeddings. 

\subsection{Discussion}

Our results yield several practical implications for CTI systems and the broader NLP community. First, TRACE can enable interested stakeholders with a mechanism to quickly understand what exploits are targeting their industry. Second, across all three model families, the single most significant factor is the choice of embedding. Corpus-trained FastText embeddings outperform off-the-shelf GloVe by 10-19\% in F1 across identical architectures. CySecBERT outperforms BERT by 5.16\% in F1 despite sharing the same architecture. This pattern echoes findings in dark web NLP~\cite{jin2022darkweb}, where general-purpose models struggle with domain-specific jargon. It also aligns with the broader observation that cybersecurity domain adaptation is essential for specialized technical language~\cite{aghaei2022securebert, park2023ctibert}.

Second, our temporal OOD evaluation, motivated by work on distribution shift~\cite{guo2023temporal, yao2022wildtime}, provides a realistic assessment choice for this domain. Exploit language evolves as new software is released, vulnerabilities are disclosed, and attack techniques shift. A model that memorizes 2019-era exploit patterns (e.g., Flash Player exploits) provides little value when deployed in 2025. Our temporal split ensures that reported metrics reflect genuine forward-looking generalization.

However, we our approach has limitations. Our gazetteer labeling, while high-precision, is imperfect. Vendors not in the gazetteer are missed, and the 7-category taxonomy aggregates organizations into broad industry groups. The Database and Gaming categories have very few test samples (2 and 30, respectively), limiting the reliability of per-category metrics for these classes. Further, our vendor-conditioned contrastive objective assumes vendor identities are meaningful groupings. Alternative metadata signals, such as project families or product families, may yield further improvements.
\section{Conclusion}

We presented TRACE, a vendor-conditioned contrastive learning framework for predicting the type of organization targeted by hacker-forum exploits. Leveraging a corpus of 8.7 million posts from 35 sources, we constructed a 129,126-sample gold-standard dataset. TRACE's CySecBERT encoder with supervised contrastive learning across vendor groups achieves F1=97.00\% with only 25 misclassifications out of 1,851 test samples, outperforming all 17 benchmarks in the temporal OOD evaluation. Our analysis revealed that the choice of embedding is critical. This model has implications for stakeholders, enabling them to quickly identify new exploits targeting their industry. Future work should deploy TRACE across the full 8.7 million-post corpus to generate large-scale temporal intelligence on shifting organizational targeting patterns, integrate CVE linkage to connect exploit discussions to specific vulnerabilities, and extend to multilingual hacker communities.

\end{document}